\pdfoutput=1
\documentclass[journal]{IEEEtran}
\usepackage{flushend}
\usepackage[table]{xcolor}
\usepackage{mathtools}
\usepackage[utf8]{inputenc}
\usepackage{times,amsmath}
\usepackage{amsfonts}
\usepackage{pstool}
\usepackage{subcaption}
\usepackage{multirow}
\usepackage{multicol}
\usepackage{enumerate}
\usepackage{graphicx}
\usepackage{mathtools, cuted}
\usepackage{MnSymbol}
\usepackage{hyperref}
\usepackage{stfloats}
\usepackage{diagbox}
\usepackage{slashbox}
\usepackage{algorithm,algorithmic}
\usepackage{bigints}
\usepackage{amsmath}
\usepackage{cite}
\begin{document}

\title{A low-cost PPG sensor-based empirical study on healthy aging based on changes in PPG morphology}

\author{\IEEEauthorblockN{Muhammad Saran~Khalid$^{\perp,}$\IEEEauthorrefmark{1}, Ikramah Shahid~Quraishi$^{\perp,}$\IEEEauthorrefmark{1},
Hadia~Sajjad\IEEEauthorrefmark{1},Hira~Yaseen\IEEEauthorrefmark{1}, \\
Ahsan~Mehmood\IEEEauthorrefmark{1}, Muhammad~Mahboob~Ur~Rahman\IEEEauthorrefmark{1}, and~Qammer~H.~Abbasi\IEEEauthorrefmark{2}}
\\
\IEEEauthorblockA{\IEEEauthorrefmark{1}Electrical engineering department, Information Technology University, Lahore, Pakistan\\}
\IEEEauthorblockA{\IEEEauthorrefmark{2}James Watt School of Engineering, University of Glasgow, Glasgow, UK\\}

\thanks{Corresponding author: Muhammad Mahboob Ur Rahman. \\ (e-mail: mahboob.rahman@itu.edu.pk).\protect\\
$^\perp$Muhammad Saran Khalid and Ikramah Shahid Quraishi are both first authors due to their equal contribution.
}
}

\maketitle


\begin{abstract}
We present the findings of an experimental study whereby we correlate the changes in the morphology of the photoplethysmography (PPG) signal to healthy aging. Under this pretext, we estimate the biological age of a person as well as the age group he/she belongs to, using the PPG data that we collect via a non-invasive low-cost MAX30102 PPG sensor. Specifically, we collect raw infrared PPG data from the finger-tip of 179 apparently healthy subjects, aged 3-65 years. In addition, we record the following metadata of each subject: age, gender, height, weight, family history of cardiac disease, smoking history, vitals (heart rate and SpO2). We pre-process the raw PPG data to remove noise, artifacts, and baseline wander. We then construct 60 features based upon the first four PPG derivatives, the so-called VPG, APG, JPG, and SPG signals, and the demographic features. We then do correlation-based feature-ranking (which retains 26 most important features), followed by Gaussian noise-based data augmentation (which results in 15-fold increase in the size of our dataset). Finally, we feed the feature set to three machine learning classifiers (logistic regression, decision tree, random forest), and two shallow neural networks: a feedforward neural network (FFNN) and a convolutional neural network (CNN). For the age group classification, the shallow FFNN performs the best with 98\% accuracy for binary classification (3-15 years vs. 15+ years), and 97\% accuracy for three-class classification (3-12 years, 13-30 years, 30+ years). {For biological age prediction, the shallow FFNN again performs the best with a mean absolute error (MAE) of 1.64. }

\end{abstract}

\section{Introduction}


Healthy aging, though most often studied in the narrow context of elderly population, is a profound phenomenon which is relevant for all age groups, i.e., children, young, and middle-age people. Healthy aging has great implications for public policy and government initiatives as it could lead to an improved quality of life, reduce the healthcare burden, and help identify the chronic diseases early \cite{darnton1995healthy}. Healthy aging is a complex, multi-faceted phenomenon that depends upon a number of factors such as genetics, physical activity, diet, smoking, alcohol consumption, diet, obesity and more. Therefore, researchers have looked into the interplay between the healthy aging and a large number of factors to date, e.g., genetics \cite{brooks2013genetics}, preventive medicine \cite{kaeberlein2015healthy}, body composition \cite{baumgartner2000body}, behavioral determinants \cite{peel2005behavioral}, nutrition and proteins \cite{evans1997nutrition}, \cite{paddon2015protein}, physical activity \cite{eckstrom2020physical}, stem cells \cite{goodell2015stem}, population-specific studies \cite{santamaria2023factors}, \cite{smith2014healthy}, \cite{guralnik1989predictors}, etc. 


A large body of literature on healthy aging aims to search for clinically viable biomarkers that could help the researchers differentiate between the healthy and unhealthy aging, for any control population group. Such biomarkers could be broadly classified as either genetic markers \cite{erikson2016whole}, \cite{obas2018aging}, or physiological markers, e.g., electrocardiogram (ECG), photoplethysmography (PPG), phonocardiogram (PCG) \cite{folkow1993physiology}, electroencephalography (EEG), magnetic resonance imaging (MRI) \cite{rossini2007clinical}, etc. This work focuses on the latter. More specifically, this work is inline with the works that focus on heart-related biomarkers (e.g., ECG, PPG, PCG, etc.), and are able to infer about the age due to their capability of capturing the physiology and/or pathology of the heart in a fine-grained manner \cite{pugh2001clinical}. 

Next, the works that utilize the biomarkers of the heart for inference about the healthy aging could be divided into two main categories: invasive methods, and non-invasive methods. The invasive methods are the traditional gold standard methods which utilize catheters and pressure sensors that are inserted into the arteries with the aim to measure arterial compliance, central aortic blood pressure, augmentation index, and more \cite{horvath2010invasive}. Nevertheless, the limitations of the invasive methods (e.g., catheters are inconvenient, costly, required trained professionals, etc.) have motivated researchers to design alternative low-cost non-invasive methods. Some popular methods include: pulse wave velocity (PWV) method, pulse transit time (PTT) method, and pulse wave analysis (PWA) method \cite{bikia2019noninvasive}---which utilize one or more of the heart-sensing modalities, e.g., ECG, PPG and PCG, etc\footnote{PWV method measures the speed at which the pressure wave travels along the arterial tree, while the PTT method measures the interval between the two points on arterial tree. PWA method, on the other hand, infers indices about the arterial stiffness from the shape of the arterial pulse signal. }. Since this paper proposes a PPG-based low-cost, non-invasive method to study healthy aging, we further narrow down our discussion to non-invasive methods for healthy aging in the rest of this paper. 

Among the non-invasive methods that study healthy aging through the lens of heart physiology, another important distinction exists. That is, there are works that aim to estimate the biological age (also known as physical age or chronological age) of a person {\cite{ladejobi202112}, and the works that aim to estimate the vascular age (also known as heart age or ECG age) of a person \cite{charlton2022assessing}, {\cite{dall2020prediction}. However, a detailed inspection of the related work reveals that the distinction between the biological age and vascular age is frequently quite blur. This is mainly due to interdisciplinary nature of the problem, and due to the fact that it is an established practice in the literature on vascular age estimation to consider biological age as a proxy for the vascular age \cite{dall2020prediction}. Thus, though the focus of this work is on biological age estimation using PPG, we use the two terminologies, i.e., biological age and vascular age interchangeably, in the rest of the paper.  

Since biological age and vascular age are tightly coupled with each other (except for the elderly population), it is imperative to briefly discuss the mechanisms that impact the vascular age of a person. The idea of vascular age begins with the observation that the structure and the function of the blood vessels undergoes a natural deterioration due to aging which becomes evident in the PPG and ECG signals \cite{charlton2022assessing}. Further, the aging process lengthens the proximal aorta, increases the stiffness and diameter of the bigger arteries, which could eventually have harmful impact on the performance of heart and other organs. This leads to a discrepancy between the biological age and vascular age (age of the heart and blood vessels) as a person ages. This discrepancy between the biological age and vascular age is a matter of great concern. More this offset is, more a person is at a risk of cardiovascular diseases (CVD), e.g., stroke, heart attack, etc. \cite{groenewegen2016vascular}. Thus, for clinical intervention, it is of utmost importance to identify people whose vascular age is higher than their physical age. 


\vspace{-0.2cm}
\subsection{Contributions} 

This work proposes to utilize a low-cost PPG sensor to design a non-invasive method for monitoring healthy aging. This problem is feasible due to the fact that biological (and vascular) aging impacts the morphology (i.e., shape and timing) of the PPG signal. The main contribution of this work is three-fold: 
\begin{itemize}
\item We construct a labelled PPG dataset using a low-cost PPG sensor-based data acquisition module to study the changes in PPG morphology due to healthy aging. Our custom dataset covers quite a wide age range of 3-65 years, and thus, stands out by covering the rare age range of 3-18 years consisting of teenagers, pre-teens, and children. Further, our custom dataset provides ethnically unique PPG data of South Asian population which is known for its distinct cardiac performance during the aging process \cite{patel2021quantifying}, \cite{pursnani2020south}. 
\item We validate the hypothesis of PPG being a viable biomarker of healthy aging, by doing two kinds of age group classification on our custom dataset using a number of machine learning (ML) and deep learning (DL) models. 1) {\it Binary classification} whereby class 0 represents the subjects in the age range of 3-15 years, while class 1 represents the subjects with age greater than 15 years. 2) {\it Three-class classification} whereby class 0 represents the subjects in the age range of 3-12 years, class 1 represents the subjects in the age range of 13-30 years, class 2 represents the subjects with age greater than 30 years. Among all AI models, the shallow feedforward neural network (FFNN) performs the best with 98\% accuracy for binary classification, and 97\% accuracy for three-class classification. 
\item We quantify the correlation between changes in PPG morphology and healthy aging, by implementing a number of regression models for biological age estimation. Our custom FFNN achieves the best accuracy with an MAE of 1.64---which is remarkable, keeping in mind that our custom dataset covers a very wide age range of 3-65 years. This again validates the hypothesis that the PPG signal contains rich and sophisticated information about the aging phenomenon. 
\end{itemize}

\vspace{-0.2cm}
\subsection{Outline} 
The rest of this paper is organized as follows. Section II summarizes selected related work on biological and vascular age estimation. Section III discusses the particulars of our data collection campaign, as well as the key steps in our data pre-processing pipeline. Section IV elaborates on the feature extraction process that utilizes the PPG signal and its derivative waveforms. Section V discusses the architecture and settings of the ML and DL models we have implemented. Section VI provides a detailed discussion of the results on both age group classification and biological age estimation. Section VII concludes the paper.
\section{Related Work}

As mentioned earlier, the distinction between the biological age and vascular age is quite blur in the literature. Therefore, this section summarizes selected related works on both problems, i.e., the works that aim to estimate the biological age, and the works that aim to estimate the vascular age.


{\it Vascular age and biological age prediction using PPG signal:}
Charlton et al.  provide a comparison of three different methods to assess hemodynamics that can help in predicting vascular age using single PPG wave, multiple PPG waves, and a combination of PPG with pulse arrival time signal. The authors concluded that shape and timing of PPG can be used to accurately measure blood flow, pulse rate, cardiac output, and vascular age despite the wave being collected by various devices and different extracted features being used \cite{charlton2022assessing}. Dall et al. obtained PPG signal of 3612 individuals aged between 18 to 79 years from Heart for Heart database and extracted 38 hand-crafted features from it to predict vascular age using ridge regression and a CNN model \cite{dall2020prediction}. Park et al. segmented PPG signal of 757 people for each beat and extracted 78 features to predict vascular age using an artificial neural network \cite{park2022vascular} . Shin et al. collected a PPG of 752 adults aged between 20-89 years and extracted hand-crafted features from it to estimate vascular age using a convolutional neural network (CNN) \cite{shin2022photoplethysmogram}. Authors of \cite{saqibanaspaperours} construct a custom 1-lead ECG dataset from subjects in the age range of 18-30 years, and exploit the distinct ECG signatures in order to train and number of ML and DL models that reliably predict vascular age, smoking habits, and gender of a person.

{\it Vascular age and biological age prediction using ECG signal:}
Ladejobi et al. collected a 12-lead ECG dataset of 25144 subjects and predicted the biological age using a CNN \cite{ladejobi202112}. Hirota et al. developed an AI-enabled sinus rhythm ECG \cite{suzuki2016nine} of 17,042 subjects aged between 20 to 90 years, to predict chronological age with a CNN \cite{hirota2023cardiovascular}. Toya et al. investigated 531 individuals for 12-lead ECG and peripheral microvascular endothelial function assessment to predict vascular aging using artificial intelligence  \cite{toya2021vascular}. Chang et al. collected ECG data from 71,741 individuals aged between 20 to 80 years to predict heart age using a CNN. The authors also presented kappa value for the multi-group classification with the age groups of $<35$, 35-49, 50-64, and $>75$ years old \cite{chang2022electrocardiogram}. 

{\it Vascular age Prediction using PPG and ECG signals:}
Chiarelli et al. collected a dataset of a multisite PPG and single lead ECG signal on 25 individuals aged between 20 to 70 years and used it as an input to extract deep features and predicted vascular age using a CNN \cite{chiarelli2019data}. 


{\it Heart age prediction via traditional score-based methods:} 
On a side note, traditional score based methods do exist that predict the heart age based on domain knowledge information, e.g., gender, blood pressure, alcohol consumption, and HDL cholesterol. Some well-known score-based methods include: Framingham risk score (FRS) \cite{wilson1998prediction}, pooled cohort equations (PCE) \cite{gibbons1997acc}, and systematic vascular risk evaluation (SCORE) method. Nevertheless, such score-based methods have their own limitations, e.g., FRS (PCE) method could overestimate the CVD risk in younger (older) age groups \cite{ko2020calibration}.


\section{Data Acquisition \& Data Pre-processing}

The proposed approach utilizes following discrete steps: data acquisition, data pre-processing, feature extraction, and inference about the biological age (classification and regression) using ML/DL models (see Fig. \ref{fig:ourmethod}).

\begin{figure}
  \centering
    \includegraphics[width=3in]{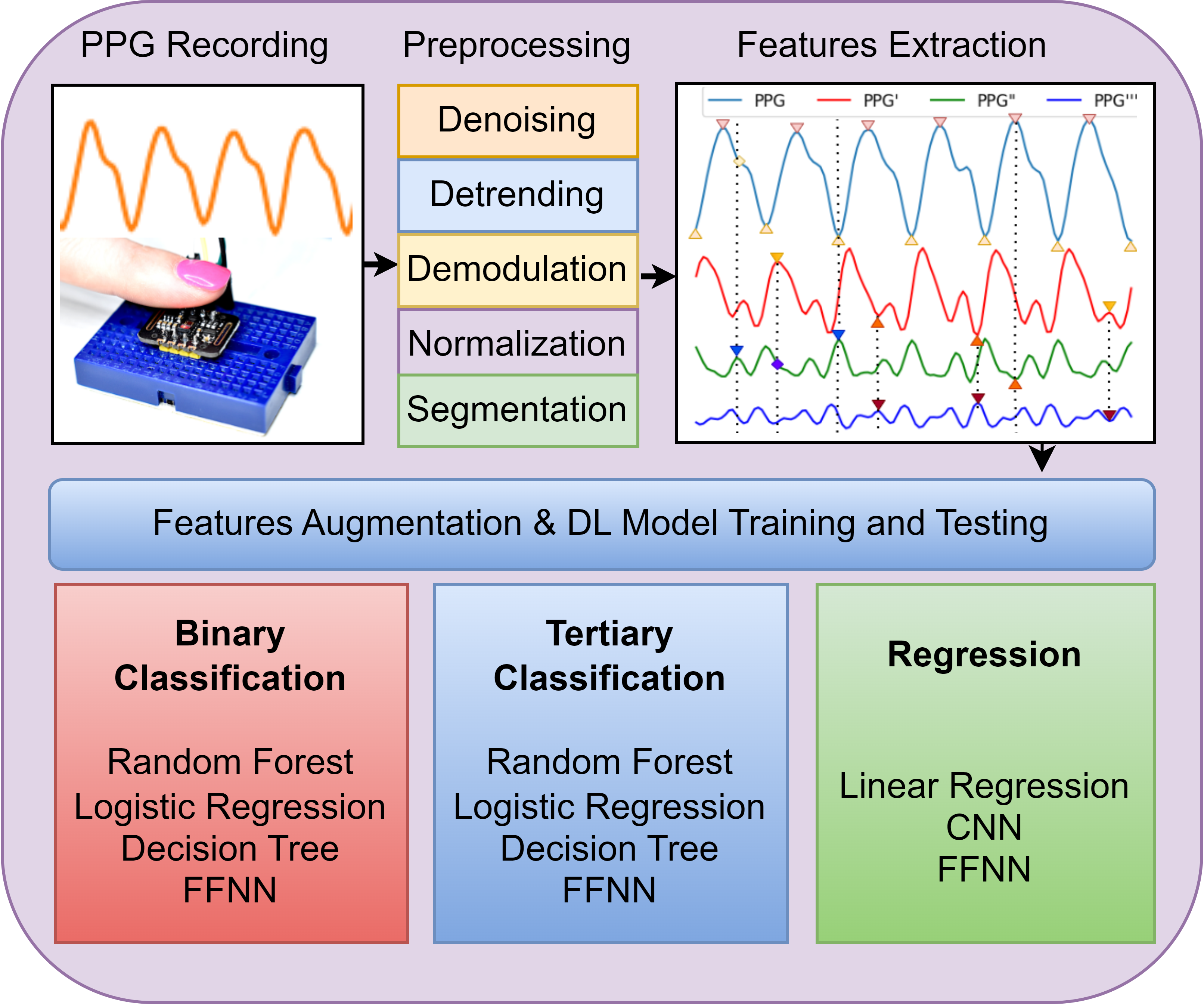}    
    \caption{Pictorial overview of our proposed approach}
    \label{fig:ourmethod}
\end{figure}

\subsection{Data Acquisition via a low-cost PPG sensor}

For the purpose of data collection, we interfaced a low-cost PPG sensor with Arduino Uno module which was in turn connected to a PC. We utilized the MAX30102 sensor (that consists of a pair of high-intensity LEDs and photo-detector) to collect raw infrared PPG data from the finger-tip of 179 healthy subjects, of age 3-65 years (114 male, 65 female). Among them, there were 45 elderly subjects, 68 young subjects, and 61 children. For each subject, we recorded the raw PPG time-series for a duration of two minutes, at a sampling rate of 400 samples/sec. This way, we collected a total of $179\times 2\times 60=21,480$ seconds worth of raw PPG data. Furthermore, for each subject, we also recorded the demographic data, e.g., age, weight, height, family history of disease, smoking history, vitals (heart rate, SpO2) etc\footnote{A permission for this study was obtained from the ethical institutional review board (EIRB) of Information Technology University (ITU), Lahore, Pakistan, before the data collection.}. 

\subsection{Data Pre-processing Pipeline}
After the data acquisition phase was complete, we performed a number of pre-processing steps on the raw PPG time-series, including the following (in order): denoising, baseline drift removal (detrending), artifact removal, demodulation, normalization, segmentation, feature extraction, and data augmentation. Let us now elaborate each pre-processing step in a bit more detail. 
\begin{enumerate}
    \item We denoised the raw PPG signals using a Chebychev Type-II filter, followed by a moving average filter. 
    \item We detrended the de-noised PPG signals by computing a centred moving average (CMA) and subtracting it from the raw signal. 
    \item The detrended and denoised PPG signal and its Hilbert transform together were used to demodulate the PPG signals (to remove the respiration induced modulation effect) as follows. The instantaneous amplitude (also known as the envelope) was extracted from the PPG signal and smoothed using a CMA filter. Then, a demodulated PPG signal was constructed by dividing the detrended signal by the smoothed envelope. 
    \item We performed $z$-score normalization on the denoised, detrended and demodulated PPG data. 
    \item We segmented the PPG signal corresponding to each subject, down to the level of a single beat. This allowed us to obtain more than 100 PPG segments (each consisting of a single heart beat) for each subject. We then averaged all the PPG segments of a subject in order to obtain a single (beat-level) noise-free PPG segment for each subject.
    \item We did feature extraction to extract a number of PPG morphological features, and appended to the feature list a number of demographic features in order to obtain a feature vector of size 60, for each of the 179 subjects. Since feature extraction is an important contribution of this work, we explain it in more detail in the next section. 
    \item We did feature ranking using the Pearson correlation method, which helped us retain the 26 most relevant features for each of the 179 subjects.
    \item Finally, we did Gaussian noise-based data augmentation in order to increase the number of examples in our dataset by 15-fold (i.e., from 179 to 2,685).  
\end{enumerate} 



\section{Feature Extraction}

Recall that the steps 1-5 of the data pre-processing pipeline allow us to collect 179 PPG segments in total, corresponding to 179 subjects (i.e., we obtain a single averaged PPG segment for each subject that is noise-free). With this, the next immediate goal is to do feature extraction from each of the 179 beat-level PPG segments. Inline with the recent works that study feature engineering for the PPG signals \cite{ppg4derivatives},\cite{dall2020prediction}, we have extracted a number of features from our PPG data: The extracted features are largely based upon: 1) various fiducial points of the PPG waveform (and intervals between them), 2) four derivative waveforms of the PPG signal\footnote{The first derivative (VPG), second derivative (APG), third derivative (JPG), and fourth derivative (SPG) of the PPG signals reflect the velocity, acceleration, jerk, and snap of the blood volume changes, respectively.}, 3) other features (e.g., standard deviation between the beats etc., see \cite{dall2020prediction}). For illustration purpose, Fig. \ref{fig:ppg4derivatives} shows few cycles of a typical PPG signal, it first four derivative waveforms and some features (corresponding to the fiducial points of the PPG signals) marked with the labels S (systolic), O (onset), N (notch). Additionally, we augmented the feature list with the demographic features, e.g., age, weight, height, family history of disease, smoking history, vitals (heart rate, SpO2) etc. In total, we managed to prepare 60 features for each of the 179 PPG segments corresponding to 179 subjects (which allowed us to execute steps 7-8 of our data pre-processing pipeline). 

\begin{figure}
  \centering
    \includegraphics[width=3in]{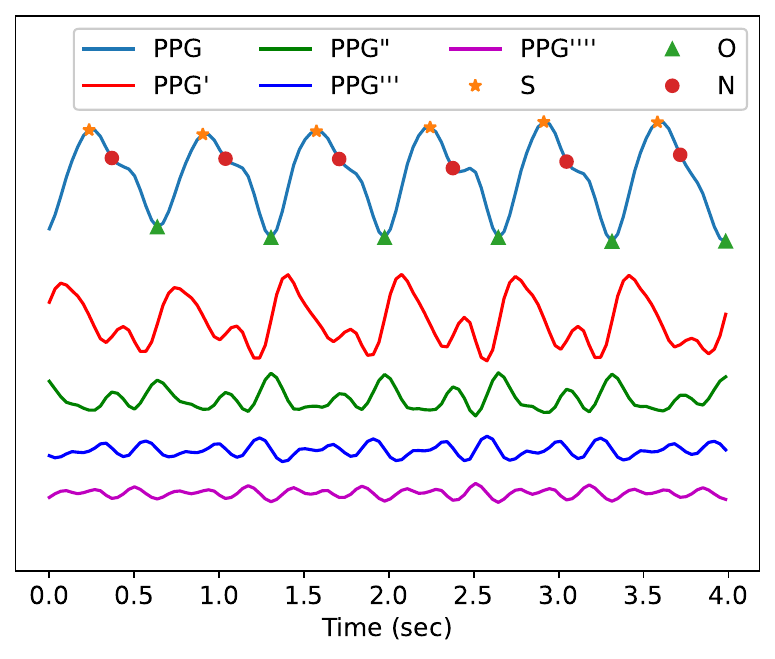}    
    \caption{A typical PPG signal and its first four derivative waveforms denoted as PPG', PPG'', PPG''', PPG'''', respectively. The features extracted from the PPG signal are marked with the labels S, O, N.
    }
    \label{fig:ppg4derivatives}
\end{figure}

\section{ML and DL Models Implemented}

Recall that this work aims to study the healthy aging phenomenon based on age-induced changes in the morphology of the PPG signal. Under this pretext, we implement a handful of ML and DL models to solve the following classification and regression problems.

{\bf Age group classification.} We do two kinds of age group classification as follows: 
\begin{itemize}
    \item {\it Binary classification} whereby class 0 represents the subjects in the age range of 3-15 years, while class 1 represents the subjects with age greater than 15 years. 
    \item {\it Three-class classification} whereby class 0 represents the subjects in the age range of 3-12 years, class 1 represents the subjects in the age range of 13-30 years, class 2 represents the subjects with age greater than 30 years. 
\end{itemize}
For both classification problems as defined above, age brackets were skillfully chosen such that the dataset remains balanced. 

We then pass the augmented dataset consisting of 2,685 examples with each example consisting of a feature vector of size $26$ as input to a number of ML and DL models (using Jupytor and Pytorch framework). As for the train-validation-test splitting, we utilize 70\% of data for training, 15\% of the data for the validation, and remaining 15\% of the data for testing. We have implemented the following ML models: Logistic Regression (LR), Decision Tree (DT), and Random Forest (RF). Additionally, we have implemented the following DL models: a feedforward neural network (FFNN), and a convolutional neural network (CNN). We used the categorical cross entropy as a loss function to train the FFNN and CNN models. We investigated various configurations of the FFNN and the CNN (by varying the number of hidden layers, and the hyperparameters). The architectures of the best-performing FFNN and CNN models are summarized in Table \ref{tab:ffnn} and Table \ref{tab:cnn}, respectively. We learn from Table \ref{tab:ffnn} that our FFNN is a shallow NN with two hidden layers only. Similarly, Table \ref{tab:cnn} tells that our custom CNN is also relatively shallow with 4 convolutional (conv.) layers only. Further, the following holds for the CNN model: kernel initializer='lecun uniform', padding='same'; L1/L2 regularization is applied in each conv. layer. Finally, we used adam optimizer, and used a learning rate of 1e-2 and 2e-2 for the FFNN and CNN, respectively.

{\bf Biological age estimation}. To quantify the correlation between changes in PPG morphology and healthy aging, we implemented the following regression models for biological age estimation: linear regression, an FFNN (as in Table \ref{tab:ffnn}), and a CNN (as in Table \ref{tab:cnn}). We used the mean absolute error (MAE) as a loss function to train all three regression models. 







\begin{table}
    \centering

\begin{tabular}{|c| c|c|c|}
        \hline
        \bf Layer&\bf Layer Name&  \bf Size & Activation \\
        \hline
 1&Input & $f_s$ & None\\ \hline
 2&Dense & 40 & Relu\\ \hline
 3&BatchNorm & 40 & None\\ \hline
 4&Dense & 10 & Relu\\ \hline
 5&BatchNorm & 10 & None\\ \hline
 6&Output & 2/3 (Classification), 1 (regression) & Softmax/Relu\\ \hline
        
    \end{tabular}
       \caption{{The shallow FFNN Model Architecture. $f_s$ = No. of features = 26.}}
       \label{tab:ffnn}
\end{table}

\begin{table}
    \centering

\begin{tabular}{| c|c|c|}
        \hline
      \bf Layer Name&  \bf Specifications & Activation \\
        \hline
 Input & $f_s$ & None\\ \hline
Conv1D & filters = 2, kernel size = 4, dilation = 1 & 'elu'\\ \hline
Conv1D & filters = 2, kernel size = 4, dilation = 1  & 'elu'\\ \hline
Dropout & Dropout rate = 0.2 & -\\ \hline
Conv1D & filters = 2, kernel size = 4, dilation = 1  & 'elu'\\ \hline
Dropout & Dropout rate = 0.2 & -\\ \hline
Conv1D & filters = 2, kernel size = 4, dilation = 1  & 'elu'\\ \hline
Dropout & Dropout rate = 0.2 & -\\ \hline
Flatten & - & -\\ \hline
Output & 2/3 (Classification), 1 (regression) & Softmax/relu\\ \hline
        
    \end{tabular}
       \caption{{The shallow CNN Model Architecture.}}
       \label{tab:cnn}
\end{table}








\section{Results}

We first discuss the results on age group classification (both binary classification and three-class classification), followed by the results on biological age estimation. We utilize confusion matrices and mean absolute error (MAE) as performance metrics for the classification problem and regression problem, respectively. Additionally, we also provide results on area under the curve (AUC) achieved by our ML and DL models, and progression of the loss function against the number of epochs during the training and validation phase, to showcase the efficacy of our DL models.  


\subsection{Promise of PPG for age group classification}

\begin{table}[h!]
\centering
\begin{tabular}{|c| c| c| c|} 
 \hline

Model	&Actual /Predicted&	(3-15) yrs	& 15+ yrs
\\

\hline
 Logistic    &	(3-15) years &	\cellcolor{blue!25}43.28\% &                      56.71\%  \\
	Regression & 15+ years &	                    	 6.17\%   &	                    \cellcolor{blue!25}93.82\%     \\
     
\hline
     Decision &	(3-15) years  &	\cellcolor{blue!25}86.94\% &                           13.05\%  \\
	Tree & 15+ years &	                          5.16\% & \cellcolor{blue!25}94.83\%  \\

\hline
     	Random & (3-15) years  &	\cellcolor{blue!25}91.41\% &                      8.58\% \\
	Forrest & 15+ years &	                    0.37\% &	\cellcolor{blue!25}99.62\% \\
     
\hline
     	CNN & (3-15) years &	\cellcolor{blue!25}22.54\% &                      77.45\% \\
	 & 15+ years &	                    5.38\% &	\cellcolor{blue!25}94.61\% 	                     \\
     
\hline
     	FFNN & (3-15) years  &	\cellcolor{blue!25}100\%  &	                   0.\% \\
	 & 15+ years &	                     0.67\% &	\cellcolor{blue!25}99.32\%  \\

 \hline
\end{tabular}
\caption{Confusion matrices of our ML and DL classifiers for the binary age group classification problem}
\label{table:2classes}
\end{table}

We first discuss the results for the binary classification problem (i.e., age being 3-15 years vs. 15+ years). Table \ref{table:2classes} outlines the confusion matrices for the three ML classifiers and the two neural networks, and thus, provides a comprehensive performance comparison of all the models we have implemented. Among the three ML models, Logistic Regression performs the worst as it records very low true positive rate (TPR) for the class 0 (i.e., age being 3-15 years). This clearly points to the fact that there is a non-linear relationship between the features extracted from the PPG waveform and the biological age, and thus, LR is not able to separate the two classes efficiently. Decision Tree, on the other hand, performs very well with a TPR of 86.94\% for class 0 and 94.83\% for class 1. Finally, Random Forest, being an ensemble tree-based method, outperforms both LR and DT methods by providing a TPR of 91.41\% for class 0 and 99.62\% for class 1. Next, the DL models. The CNN model remains biased towards the class 1 (i.e., age being 15+years) as it records a very low TPR for the class 0. This poor performance of the CNN model could be due to the small size of our custom dataset. The FFNN model, on the other hand, outperforms the CNN model as well as the three ML models by achieving a 100\% TPR for class 0 and 99.2\% TPR for class 1.

\begin{table}[h!]
\centering
\begin{tabular}{|c| c| c| c|c|} 
 \hline

Model	&Actual /Predicted&	3-12 yrs &	13-30 yrs	& 30+ yrs
\\

\hline
 Logistic    &	3-12 years  &	\cellcolor{blue!25}76.52\% &                      11.03\% &	                   12.44\%  \\
	Regression & 13-30 years &	                    18.80\%    &	\cellcolor{blue!25}73.29\%   &	                    7.90\%     \\
    & 30+ years &	                   35.68\%  &	                  5.57\% &	\cellcolor{blue!25}58.73\% \\
 
\hline
     Decision &	3-12 years  &	\cellcolor{blue!25}83.33\% &                           8.21\% &	                   8.21\%  \\
	Tree & 13-30 years &	                           8.17\% & \cellcolor{blue!25}87.46\% &	                          4.35\% \\
    & 30+ years &	 11.52\% &                           7.06\% &	\cellcolor{blue!25}81.41\% \\
     
\hline
     	Random & 3-12 years &	\cellcolor{blue!25}99.76\% &                      0.23\% &	                   0\% \\
	Forrest & 13-30 years &	                    4.90\% &	\cellcolor{blue!25}95.09\% &	                    0\% \\
     & 30+ years &	                   6.694.22\% &	                      0.74\% &	\cellcolor{blue!25}92.56\% \\
\hline
     	CNN & 3-12 years  &	\cellcolor{blue!25}74.05\% &                      17.08 &	                   8.86\% \\
	 & 13-30 years &	                    17.85\% &	\cellcolor{blue!25}74.28\% &	                     7.85\% \\
     & 30+ years &	                   39.60\% &	                      5.94\% &	\cellcolor{blue!25}54.45\% \\
\hline
     	FFNN & 3-12 years  &	\cellcolor{blue!25}98.73\% &                       0.63\% &	                   0.63\% \\
	 & 13-30 years &	                    0\% &	\cellcolor{blue!25}98.57\% &	                    1.42\% \\
     & 30+ years &	                   0\% &	                      0\% &	\cellcolor{blue!25}100\% \\

 \hline
\end{tabular}
\caption{Confusion matrices of our ML and DL classifiers for the three-class age group classification problem}
\label{table:3classes}
\end{table}

Next, we summarize the results for the three-class classification problem (i.e., age being 3-12 years, or, 13-30 years, or 30+ years) by providing the confusion matrices of all the three ML classifiers and the two neural networks in Table \ref{table:3classes}. Among the three ML models, Logistic Regression again performs the worst, due to the fact that the decision boundary between the classes is highly non-linear. This premise is proven by the Decision Tree method which records a TPR of 83.33\% for class 0, 87.46\% for class 1, and 81.41\% for class 2. Finally, Random Forest, being an ensemble tree-based method, again outperforms both LR and DT methods by providing a TPR of 99.76\% for class 0, 95.09\% for class 1, and 92.56\% for class 2. As for the DL models, the CNN model records a mediocre performance in terms of TPR for class 0 and class 1, and very poor performance for class 2. Again, this poor performance by the CNN is probably due to small size of our custom dataset (even after data augmentation by a factor of 15). The FFNN model again outperforms the CNN model and the three ML models by achieving a 98.73\% TPR for class 0, 98.57\% TPR for class 1, and a 100\% TPR for class 2.

\begin{table*}
    \scriptsize
    \centering
    \setlength\tabcolsep{1pt}
  \begin{tabular}{|c|c|c|c|c|c|c|c|c|c|c|c|}
        \hline
        \bf Method & \multicolumn{2}{|c|}{\textbf{Logistic Regression}}  & \multicolumn{2}{|c|}{\textbf{Decision Tree }} & \multicolumn{2}{|c|}{\textbf{ Random Forest}} & \multicolumn{2}{|c|}{\textbf{CNN}}  & \multicolumn{2}{|c|}{\textbf{FFNN }} \\\hline
        \hline
        Performance Metrics&AUC&Accuracy& AUC &Accuracy&AUC&Accuracy&AUC&Accuracy&AUC&Accuracy\\\hline
        \bf Binary Classification& 0.67&0.80& 0.91& 0.93&0.95&0.97&0.73&0.83& \bf 0.98& \ \bf 0.98\\ \hline
        \bf Three-class  Classification& 0.75&0.69& 0.92& 0.89&0.96&0.95&0.74&0.67& \bf 0.98& \bf 0.97\\ \hline

    \end{tabular}
        \caption{{AUC and accuracy results of our ML and DL models for binary and three-class age group classification}}
       \label{tab:acc_compare_classify}
\end{table*}

Fig. \ref{fig:epoch_plot_classify} provides insights into the progression of the performance of our FFNN and CNN models against the number of epochs, during the training and the validation phase. Specifically, Fig. \ref{fig:epoch_plot_classify} (a)-(c) provide a vivid performance comparison between the FFNN and CNN for binary age group classification. First, we observe that the loss functions for both FFNN and CNN reach the plateau in about 200 epochs. We further learn from Fig. \ref{fig:epoch_plot_classify} (a)-(c) that the CNN saturates at inferior values of the loss function, AUC and the accuracy compared to the FFNN. In other words, the CNN exhibits a kind of underfitting when compared with the FFNN, which could be due to the small size of our custom dataset. Next, Fig. \ref{fig:epoch_plot_classify} (d)-(f) illustrate the performance comparison between the two DL models in terms of loss function, accuracy and AUC, but this time for the three-class age group classification problem. We observe from Fig. \ref{fig:epoch_plot_classify} (d)-(f) that the phenomenon of so-called underfitting of the CNN compared to the FFNN becomes more pronounced for three-class classification.

Table \ref{tab:acc_compare_classify} summarizes the overall accuracy as well as the AUC of the three ML and the two DL classifiers, for both binary and three-class age group classification problem. We make the following observations. 1) The Random Forest method outperforms the other two ML classifiers with an overall accuracy of 97\% and AUC of 0.95 for binary classification, and an accuracy of 95\% and AUC of 0.96 for three-class classification. 2) The FFNN outperforms all ML classifiers as well as the CNN with an accuracy of 98\% and AUC of 0.98 for binary classification, and an accuracy of 97\% and AUC of 0.98 for three-class classification. 

Overall, a high age group classification accuracy achieved by the Random forest method and the FFNN model (more than 95\%) as depicted in Tables \ref{table:2classes}-\ref{tab:acc_compare_classify} attests to the fact that {\it PPG signal is indeed a reliable biomarker for analyzing healthy aging (i.e., the morphology of the PPG signal changes significantly as the age progression takes place)}.

\begin{figure*}
     \centering
     \begin{subfigure}[b]{0.3\textwidth}
         \centering
         \includegraphics[width=\textwidth]{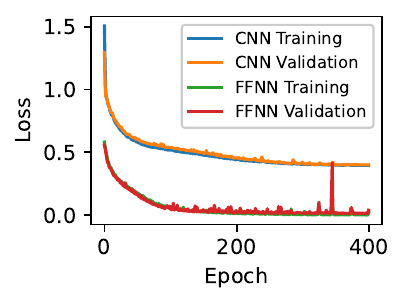
         }
         \caption{Training and validation loss curves for binary classification}
     \end{subfigure}
     \hfill
     \begin{subfigure}[b]{0.3\textwidth}
         \centering
         \includegraphics[width=\textwidth]{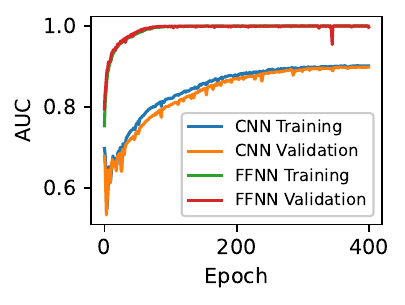}
         \caption{Training and validation AUC curves for binary classification.}
     \end{subfigure}
     \hfill
     \begin{subfigure}[b]{0.3\textwidth}
         \centering
         \includegraphics[width=\textwidth]{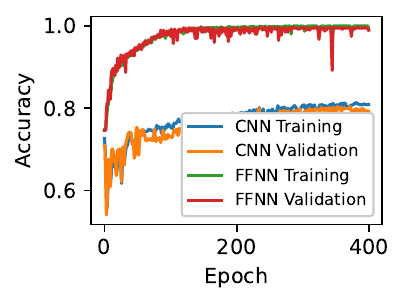}
         \caption{Training and validation accuracy curves for binary classification}
     \end{subfigure}
     \hfill 
     \begin{subfigure}[b]{0.3\textwidth}
         \centering
         \includegraphics[width=\textwidth]{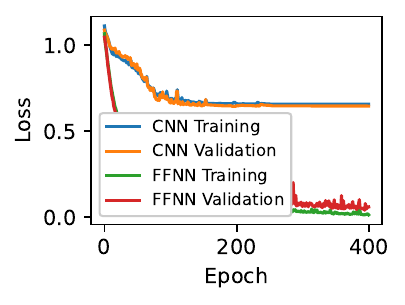}
         \caption{Training and validation loss curves for three-class classification}
     \end{subfigure}
     \hfill
     \begin{subfigure}[b]{0.3\textwidth}
         \centering
         \includegraphics[width=\textwidth]{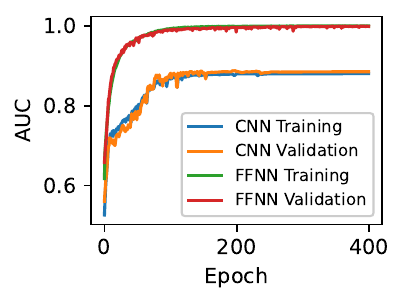}
         \caption{Training and validation AUC curves for three-class classification}
     \end{subfigure}
     \hfill
     \begin{subfigure}[b]{0.3\textwidth}
         \centering
         \includegraphics[width=\textwidth]{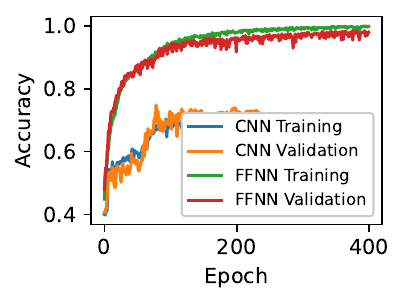}
         \caption{Training and validation accuracy curves for three-class classification}
     \end{subfigure}
      
        \caption{{Progression of performance of the CNN and the FFNN during the training and validation phase, for binary and three-class age group classification.}}
        \label{fig:epoch_plot_classify}
\end{figure*}

\begin{figure*}
     \centering
     \begin{subfigure}[b]{0.3\textwidth}
         \centering
         \includegraphics[width=\textwidth]{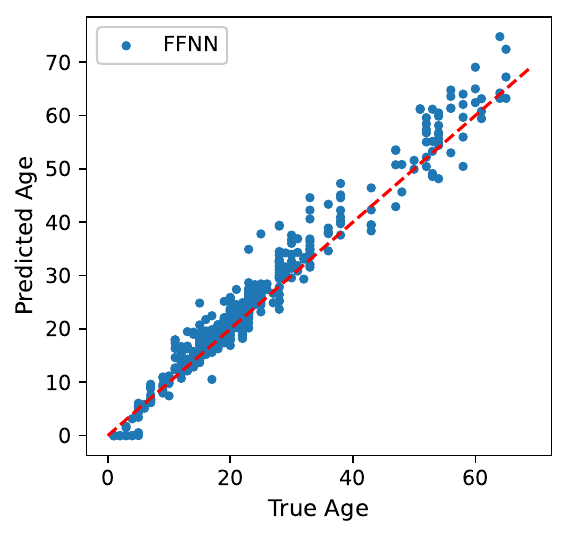}
         \caption{FFNN regression, MAE = 1.64}
     \end{subfigure}
     \hfill
     \begin{subfigure}[b]{0.3\textwidth}
         \centering
         \includegraphics[width=\textwidth]{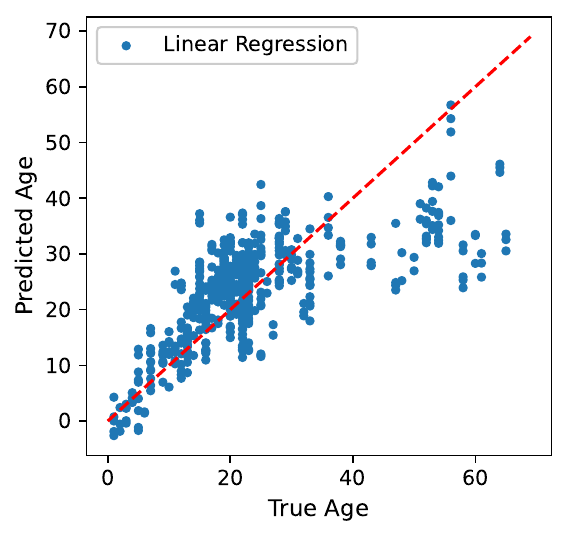}
         \caption{Linear regression, MAE = 7.04}
     \end{subfigure}
     \hfill
     \begin{subfigure}[b]{0.3\textwidth}
         \centering
         \includegraphics[width=\textwidth]{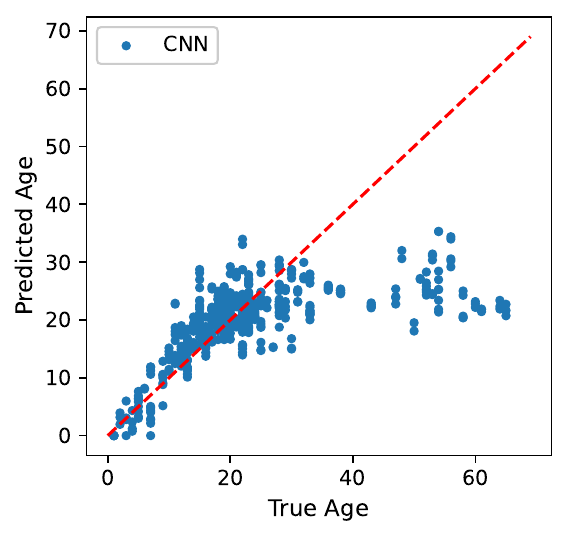}
         \caption{CNN regression, MAE = 6.30}
     \end{subfigure}
      \hfill
    
        \caption{{Performance of the FFNN, CNN and linear regressor for biological age prediction.}}
        \label{fig:regression_mae}
\end{figure*}

\begin{figure}[!ht]
  \centering
    \includegraphics[width=3in]{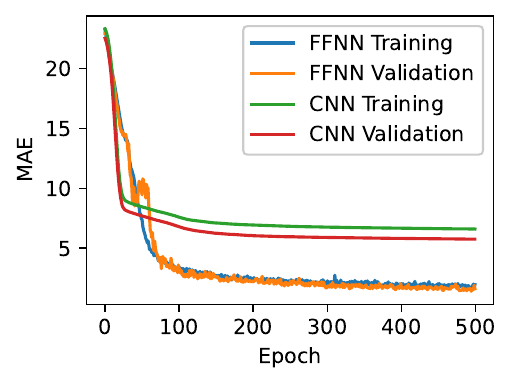}
    \caption{Progression of MAE loss function of the CNN and the FFNN during the training and validation phase, for biological age prediction problem.}
    \label{fig:epoch_plot_regression}
\end{figure}

\subsection{Promise of PPG for biological age estimation} 

Fig. \ref{fig:regression_mae} illustrates the performance of the three regression methods, i.e., the FFNN, the CNN and the linear regressor, by means of a scatter plot. Following comments are in order. 1) The FFNN stands out with an MAE of 1.64 years, while the CNN and linear regressor have relatively large MAE of 6.3 years and 7.04 years, respectively. 2) A closer look at Fig. \ref{fig:regression_mae} reveals that both CNN and linear regressor have a bias towards the younger age brackets. That is why, both models under-predict the age of the middle-age and elderly people (with age 35+ years) to predict them as relatively younger people. 3) This implies that the FFNN is a good estimator of biological age of a person over a very wide age range, i.e., 3-65 years, while the CNN and the linear regressor could reliably output the biological age over a narrow range of 3-40 years only. 4) We observe that the FFNN slightly overestimates the biological age of very elderly people in the age range of 60-65 years. This points to the fact the FFNN model is also a good predictor of the vascular age of a person. That is, the FFNN internally captures the relevant features in the PPG signal which contain rich information about the health of the cardiovascular system of an elderly person.

Fig. \ref{fig:epoch_plot_regression} shows the decay in the MAE loss function for both the FFNN and the CNN models during the training and the validation phase, as we increase the number of epochs. We observe that the MAE loss function reaches the plateau in about 200 epochs, for both FFNN and CNN. Again, it is quite evident that the CNN exhibits a kind of underfitting when compared with the FFNN, which is most likely due to the small size of our custom dataset.

{\it Comparison with related work:}
As we have mentioned earlier, this work is mainly a contribution on the feasibility of healthy aging analysis using a low-cost PPG sensor. To this end, we showcase the viability of the PPG signal as an efficient biomarker for healthy aging by means of age group classification. But since we have made the age brackets such that the dataset remains balanced, and that there are significant differences in the attributes and modalities of our custom dataset and other public datasets; therefore, it is unnatural to compare the performance of our work against the related work on age group classification, e.g., \cite{attia2019age}, \cite{chang2022electrocardiogram} do age group classification with custom age brackets using 12-lead ECG data. However, there has been works that either estimate the physical age of a person (also known as biological or chronological age), or the vascular age of a person (also known as ECG age or heart age). Therefore, though this was not the main focus of this work, the regression problem solved in this work could also be thought as a contribution to the set of works that do vascular aging estimation by taking the biological age as a proxy for the vascular age of a person\footnote{Note that the true vascular age of a person could only be measured in a hospital setting where trained staff carries out invasive testing to estimate various gold standard parameters, e.g., arterial stiffness, pulse transit time, pulse transmit time, etc \cite{charlton2022assessing}. That is why, it is a routine practice by the researchers to take the biological age as a proxy for the vascular age \cite{dall2020prediction}.} \cite{dall2020prediction}. With this in mind, we have constructed Table \ref{tab:compare_sota} that qualitatively compares the performance of our work with the state-of-the-art (SoTA) on biological age estimation and vascular age estimation. A detailed inspection of Table \ref{tab:compare_sota} reveals the following. 1) Our custom dataset covers quite a wide age range of 3-65 years, and stands out by covering the rare age range of 3-18 years consisting of teenagers, pre-teens, and children. 2) the MAE performance of our proposed method (pre-processing pipeline plus CNN model) is quite superior compared to the SoTA. 

Finally, keeping in mind the population-specific studies \cite{santamaria2023factors}, \cite{smith2014healthy}, \cite{guralnik1989predictors}, we further want to remind the reader that our custom dataset is a contribution to the community in itself as it provides ethnically unique PPG data of South Asian population which is known for its distinct cardiac performance during the aging process \cite{patel2021quantifying}, \cite{pursnani2020south}.

\begin{table}
    \scriptsize
    \centering
    \setlength\tabcolsep{1pt}
  \begin{tabular}{|c|c|c|c|c|c|c|c|c|c|c|}
        \hline
        Work & $\mathcal{N}$ & Age  (years)& Biosignal & AUC & MAE &RMSE& Age\\
        \hline
        Chiarelli at al. \cite{chiarelli2019data}&25&20-70& PPG \& ECG&-&-&7.00&$\checkmark \times$\\\hline
        Chang et el. \cite{chang2022electrocardiogram}&71741&20-80&ECG&-&6.89&-&$\checkmark \times$\\\hline
        Hirota et al. \cite{hirota2023cardiovascular}&17042&20-90&ECG&0.679&6.13&-&$\checkmark$\\\hline
        Ladejobi et al. \cite{ladejobi202112}&25144& $\geq$ 30 &ECG&-&8.08&-&$\checkmark$\\\hline
        Shin et al. \cite{shin2022photoplethysmogram}&752&20-89&PPG&-&8.1&10.0&$\checkmark \times$\\\hline
        Park et al. \cite{park2022vascular}&757&0-89&PPG&-&-&10.0&$\checkmark \times$\\\hline
        Dall et al. \cite{dall2020prediction}&4769&18-79&PPG&0.953&-&-&$\checkmark \times$\\\hline
        {\bf Our work} &179&3-65&PPG&-&1.64&-&$\checkmark $\\\hline

    \end{tabular}
    \caption{Qualitative comparison of our work with the SoTA that estimates biological age and/or vascular age. {$\mathcal{N}$ represents  number of subjects in the dataset. $\checkmark \times$ represents the fact that a particular work has estimated the vascular age or heart age or healthy vascular age by using the biological age as a ground truth. '$\checkmark$' represents a work that has estimated the biological age itself.} }
       \label{tab:compare_sota}
\end{table}

\section{Conclusion}

We performed an experimental study to quantify the relationship between the morphology of the PPG signal and healthy aging. We first collected the raw infrared PPG data using a non-invasive low-cost MAX30102 PPG sensor (as well as the metadata) from 179 healthy subjects, aged 3-65 years. We pre-processed the raw PPG data to remove noise, artifacts and baseline wander, extracted a number of features based upon the first four PPG derivatives, did data augmentation and fed the data to our ML and DL methods. For both age group classification and biological age estimation problems, the shallow FFNN achieved great accuracy, which attests to the fact that PPG is indeed a promising (i.e., low-cost, non-invasive) biomarker to study the aging phenomenon.

Looking forward, we note that there are a number of factors that could influence the healthy aging phenomenon, e.g., distinct physiological characteristics of individuals, different kinds of lifestyles, genetic factors, food intake, smoking and alcohol consumption, etc. Thus, AI-assisted fusion of existing datasets on vascular and biological age estimation with various modalities (e.g., PPG, ECG, and more) is needed, in order to construct large and diverse datasets on healthy aging. Further, preparation of large hybrid datasets (consisting of both real data and generative AI-based synthetic data) which capture all the aforementioned factors related to the aging process is the need of the hour. This would in turn allow researchers to develop robust, trustworthy, and generalizable AI models that could overcome the so-called challenge of {\it diverse populations} (due to genetic variation, varied lifestyles, and more) in order to output more fine-grained analytics on healthy aging.

\footnotesize{
\bibliographystyle{IEEEtran}
\bibliography{main}
}
\vfill\break
\end{document}